**Towards a new generalized space expansion dynamics applied to the rotation of galaxies and Tully Fisher law**

Jacques Fleuret

jacques.fleuret@telecom-paristech.org

Abstract

Up to now, the rotational velocities of galaxies are not clearly understood and the experimental Tully Fisher rule, linking the total galactic mass to the fourth power of the velocity, through an acceleration coefficient of about $10^{-10}$ m/s$^2$ has not found a deep theoretical explanation. Tentative proposals (MOND theory of a modified Newton's law and extraneous dark matter) do not bring a definite clarification. We propose here a new approach to this problem, without exotic matter and using the classical Newton force. But we introduce a new additional universal acceleration, which could represent a universal expansion law valid at the scale level of a galaxy. We show that this hypothesis leads to a good description of the observed variations of the galactic transverse velocity. It can be considered as a consequence of the Scale Expansion Cosmos theory (SEC) introduced by J. Masreliez, but we postulate that the space expansion acceleration universally applies at any scale. We obtain a formal derivation of the Tully Fisher law, linking the constant galactic transverse velocity to its total mass, via the universal minimum acceleration. We derive a good estimate of the TF acceleration coefficient and show that expansion should be proportional to the square root of the local volumic mass density. Our conjecture is in fact a new dynamics principle which could be applied to many other physical problems at different scales. Applying it to the range of the solar planet system confirms the well known Kepler laws, at least as a valid approximation for the order of magnitude of the solar system.

**Keywords**:

Galaxy, flat rotation curves, Tully-fisher, cosmology, expansion, dark matter

## 1 Introduction

Up to now, the rotational velocities of galaxies are not clearly understood. The gathered observations (Courteau et al., 2003; Palunas & Williams, 2000) of velocity curves do not fit well with gravitation laws. The experimental Tully Fisher rule has been established (McGaugh, 2011; Tonini et al., 2011; Mo & Mao, 2000; Verheijen, 2001) but has not found a deep theoretical foundation. It introduces an acceleration coefficient $\gamma_0$ of about $10^{-10}$ m/s$^2$ linking the total galactic mass to the fourth power of the measured transverse velocity.

Two new principles have been attempted in response: the assumption of a modified Newton's law (MOND theory) for small accelerations (McGaugh, 2002; Cardone et al., 2011) and the hypothesis of a dark halo made of baryonic or non-baryonic black matter. Despite a large number of studies of this last hypothesis, the conclusions are not clear (McGaugh, 2011; Tonini et al., 2011; Mo & Mao, 2000; Verheijen, 2001) and (Bottema & Verheijen, 2001; Bottema, 2002, Feng & Gallo, 2010, Bienaymé, 1999).

Very few papers tend to envision other approaches (Taylor, 1998; Mizony, 2003, Fuchs et al., 2004; Cooperstock & Tieu, 2007).

On the other hand, the expansion of the Universe is well established involving expansion velocities proportional to distance according to the well-known Hubble law (Riess, 1998). Few studies have tempted to apply it at galactic level (Nandra et al., 2012). In his original "Scale Expansion Cosmos" theory, J. Masreliez envisions space and time expansion, which would imply galactic flat rotation curves with spiral star trajectories (Masreliez, 2012; Masreliez, 2004a,b).

We propose here to introduce a new additional universal acceleration, which could represent a universal expansion law valid at the scale level of a galaxy. We'll show that it can be considered as a "cosmic drag" resulting from a SEC theory, where expansion is not homogeneous but depends on the local scale of space. This additional acceleration allows a good description of the observed variations of the galactic rotational curves, without exotic matter. More generally, the conjecture of a universally scaled expansion force will allow us to derive theoretical expectations for the main physical quantities linking expansion to galaxy dynamics.

**2 Initial assumption**

2.1 *The problem of the rotation of a planar mass distribution*

The rotation of a planar mass distribution submitted to Newtonian gravity follows the well-known equations, written in polar coordinates (we neglect the reciprocal influence of the star on the galaxy, due to the huge difference of masses):

$$2\dot{r}\dot{\theta} + r\ddot{\theta} = 0 \tag{1}$$

$$\ddot{r} - r\dot{\theta}^2 = -\Gamma_t(r) \tag{2}$$

Where $\Gamma_t(r)$ is the central gravitation acceleration due to the whole baryonic mass distribution in the galaxy at the time t considered (radial symmetry is supposed) and $\theta$ is the polar angle.

These equations explain the planar movement, through (1) and the resulting angular momentum $C = rv_\theta(r) = r^2\dot{\theta}$,

where the (transverse) rotational velocity is:

$$v_\theta = r\dot{\theta} \tag{3}$$

The observed velocity curves $v_\theta(r)$ for planar galaxies show a very rapidly growing part near the origin, usually followed by a slowly decreasing function which remains quasi constant for a large range of radius (Palunas & Williams, 2000; McGaugh, 2011; Tonini et al., 2011). But eq (1) necessarily leads to a transverse velocity inversely proportional to $r$. Any other additional force (collisions, thermic or electromagnetic forces, etc.) could be envisioned to contribute to the process: this would not change that last result unless the force had a transverse part, thus changing eq. (1).

2.1 *Conjecture of a universal expansion force acting at galaxy range level*

We examine here the consequences of an additional expansion acceleration which would be valid at a galaxy range scale, and follow the given <u>vectorial</u> relationship:

$$\vec{\gamma}_{exp} = \frac{\dot{r}}{r}\vec{v} \qquad (4)$$

When compared to the classical Newtonian acceleration for a quasi circular orbit, we immediately point out that it is in the ratio:

$$\left|\frac{\gamma_{exp}}{\Gamma_{Newton}}\right| = \left|\frac{\dot{r}}{v}\right| \qquad (5)$$

When acceleration (4) is added, eq. (1) is rewritten as:

$$2\dot{r}\dot{\theta} + r\ddot{\theta} = \frac{\dot{r}}{r}v_\theta \qquad (1b)$$

Then from (3):

$$\dot{r}\dot{\theta} + r\ddot{\theta} = 0 \qquad (1c)$$

which means that $v_\theta$ does not depend on time.

Note that the $\dot{r} = 0$ solutions of (1b) still require a constant angular momentum. More generally, eq (1b) does not lead to a constant angular momentum and seems to be in contradiction with the well known Kepler laws for planet motions. This point will be clarified in section 3.9. The rest of the paper is focused on the star rotations within a galaxy.

**3 Resulting developments for galaxy dynamics and expansion**

3.1 *Temporal radial evolution*

We also rewrite the radial equation as:

$$\ddot{r} - r\dot{\theta}^2 = -\Gamma_t(r) + \frac{\dot{r}}{r}\dot{r} \qquad (2b)$$

And, using (3):

$$r\ddot{r} - \dot{r}^2 = v_\theta^2 - r\Gamma_t(r) \qquad (2c)$$

this equation describes the temporal evolution of r.

In this paper, we examine the possibility of a solution with a constant transverse velocity, according to (1c):

$$v_\theta = r\dot\theta = v_0 \quad (6)$$

Mathematically, (6) is not a necessary condition, but it is a sufficient one, describing one possible solution of the dynamics equations.

Then we should have:

$$r\ddot r = v_0^2 + \dot r^2 - r\Gamma_t(r) \quad (2d)$$

which can be written as:

$$\frac{d}{dt}\left(\frac{\dot r}{r}\right) = \frac{v_0^2}{r^2} - \frac{\Gamma_t(r)}{r} \quad (2e)$$

We assume here that $\frac{\dot r}{r}$ does not depend explicitly on time: $\frac{\partial}{\partial t}\left(\frac{\dot r}{r}\right) = 0$.

For very large radii:

$$\Gamma_t(r) \cong \frac{GM_t}{r^2} \quad (7)$$

where $M_t$ is the total mass of the galaxy. Then, eq. (2e) can be written as:

$$\frac{\dot r}{r}\frac{\partial}{\partial r}\left(\frac{\dot r}{r}\right) = \frac{v_0^2}{r^3} - \frac{GM_t}{r^4} \quad (8)$$

Resulting in:

$$\left(\frac{\dot r}{r}\right)^2 = h^2 - \frac{v_0^2}{r^2} + \frac{2}{3}\frac{GM_t}{r^3} \quad (9)$$

where $h^2$ is an integration constant.

Let-us now introduce the maximum radius $r_M$ of the galaxy, where it is reasonable to suppose the two independent hypotheses:

1) expansion at $r = r_M$ is radially quasi-stationary:

$$\left[\frac{\partial}{\partial r}\left(\frac{\dot r}{r}\right)\right]_{r=r_M} \cong 0 \quad (10)$$

Then, from (8):

$$v_0^2 r_M \cong GM_t \qquad (11)$$

And for $r \cong r_M$ we obtain:

$$\left(\frac{\dot{r}}{r}\right)^2 \cong h^2 - \frac{1}{3}\omega_0^2 \qquad (12)$$

Where $\omega_0$ represents the angular velocity at $r_M$:

$$\omega_0 = \frac{v_0}{r_M} \qquad (13)$$

2) We also assume that expansion is very small at the galactic edge:

$$\left(\frac{\dot{r}}{r}\right)_{r=r_M} \ll \omega_0 \qquad (14)$$

and obtain from (12):

$$h \cong \frac{\omega_0}{\sqrt{3}} \qquad (15)$$

Since there is not only one galaxy in the universe, the constant h should not be confused with the Hubbel Constant H, which results from the influences of all galaxies with their own particular parameter values (equivalent of $v_0, r_M$ and $\omega_0$), from all directions of space.

We shall derive the value of the Hubbel Constant by another approach further down.

Finally, we obtain here quasi circular spiral trajectories and expansion depends on radius. It is minimal at the galaxy edge (eq. 14) and can be practically neglected there.

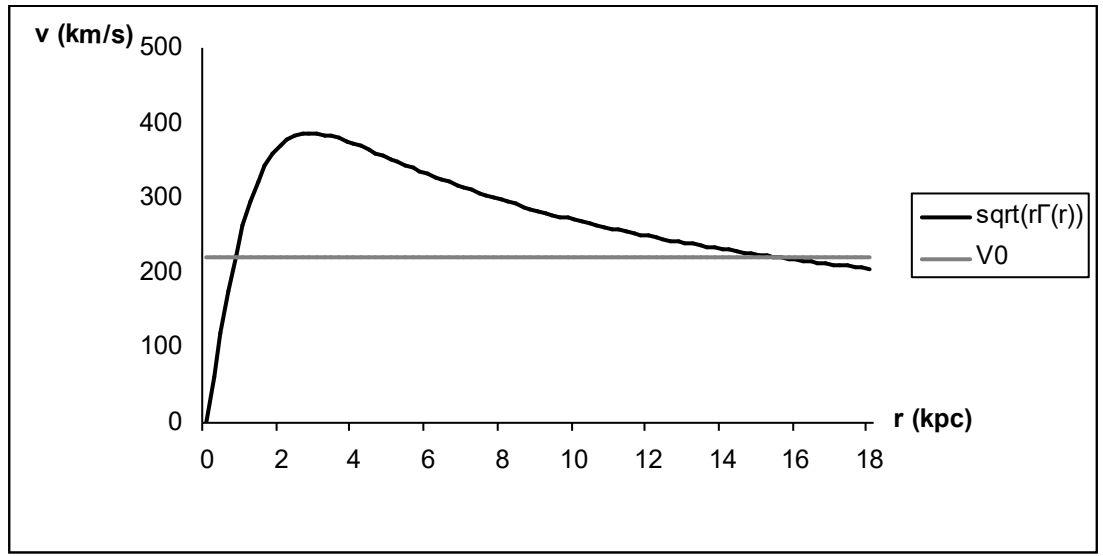

FIG.1. Illustration of the transverse velocity curve (for a Kuzmin mass density model)

3.2 *Transverse velocity curve*

Equations (1b) and (2d) illustrate the way our proposed process works.

Dark matter should produce an inward acceleration.

Our (small) outward acceleration (4) is vectorial. Its transverse part stabilizes $v_\theta(t)$ (eq. 1b). Furthermore, eq. (2d) will also tend to regulate transverse velocity.

To illustrate the process of our proposal, fig. 1 has been drawn as an instance for the case of a Kuzmin surface mass density (Binney & Tremaine, 1994).

Here, $r\Gamma(r)$ results from a surface mass density $\rho = \dfrac{r_0 M_t}{2\pi \left(r^2 + r_0^2\right)^{\frac{3}{2}}}$ :

$$r\Gamma(r) = \frac{GM_t r^2}{\left(r^2 + r_0^2\right)^{\frac{3}{2}}} \cong \frac{v_0^2 \left(\dfrac{r}{r_M}\right)^2}{\left[\left(\dfrac{r}{r_M}\right)^2 + \left(\dfrac{r_0}{r_M}\right)^2\right]^{\frac{3}{2}}}$$

The chosen parameters are $M_t = 3,5\ 10^{41}$kg, $r_0 = 2 kpc$ and $v_O = 220 km/s$.

There are two values of $r$ for which eq (2e) is null: $r_m \cong 0,7 kpc$ and $r_M \cong 15,6 kpc$.

Above $r_M$, the expansion is accelerated (eq.2e) and it is decelerated in the range $r_m \leq r \leq r_M$.

For $r \cong r_M$, we have quasi circular spirals ($\dot{r} \neq 0$ but small) and stable expansion.

Furthermore, we see that $v_\theta^2$ in eq. (2) has been replaced by $v_0^2 + \dot{r}^2$ in eq. (2c). The radial velocity term tends to regulate the radial variation of transverse velocity: along an outward spiral trajectory, $v_\theta$ should decrease according to Newton law, but the decrease of $\dot{r}^2$ due to eq. (8) allows $v_0$ to remain constant.

It must be noted that the sign of $\dot{r}$ is not specified by eq. (9): the spiral trajectories can be inward or outward. Here, $\dot{r} > 0$ have been chosen, to match with what occurs outside the galaxy (outward expansion). But contraction can also be envisioned. In this case, instead of "decelerated expansion", we could have "accelerated contraction". Only experimental observations can confirm the mathematical choice.

We also emphasize the fact that – even if our expansion term $\dfrac{\dot{r}^2}{r}$ "goes the wrong way" when compared to the hypothetical dark matter influence – it does not produce en explosion. For small $\dot{r}$, $\ddot{r}$ remains negative (eq. 2d) and smaller than the Newton acceleration.

In the case $\dot{r} \geq 0$, we have obtained outward decelerated spiral trajectories.

If $\dot{r} \leq 0$, we would have inward (slowly) accelerated contraction.

3.3 *Case of low radii and summary*

For low radii ($r \leq r_m$), experimental observations show a very rapidly growing $v_\theta(r)$ curve. In this case, the Newton acceleration can be approximated as:

$$\Gamma_t(r) \cong \frac{\pi \rho_0 r^2 G}{r^2} = \pi \rho_0 G \qquad (16)$$

where $\rho_0$ is the central (two dimensional) mass density. Assuming that eq. (2c) remains null in this range where spiral trajectories are still valid [$\frac{d}{dt}\left(\frac{\dot{r}}{r}\right) = 0$ or $\dot{r} = h_0 r$], we obtain:

$$v_\theta = \sqrt{\pi \rho_0 G r} \qquad (17)$$

which fits observation.

To resume:

Near center, "stable" exponential expansion produces the rapidly growing curve (17). Beyond $r_m$, the transverse velocity remains constant and (decelerated) expansion takes place up to $r_M$, where (small) stable expansion is observed. Outside the galaxy, expansion is accelerated.

The observed $v_\theta(r)$ curve is nothing but the $\sqrt{r\Gamma(r)}$ curve up to $r_m$ and $v_\theta(r) = v_0$ above.

3.4 *Eulerian transverse velocity*

The above description concerns individual trajectories of a single star (Lagrangian formulation).

When we consider a galaxy, the velocities of all stars probably tend to regulate themselves due to interaction processes: for medium sized stars having approximately equal masses, this process would lead to a quasi constant (Eulerian) radial velocity. The Eulerian point of view is given in Appendix, where it is shown that a constant transverse velocity remains possible, under the condition that acceleration (4) is added.

3.5 *Expansion acceleration versus SEC cosmic drag*

In the Scale Expansion Cosmos (SEC) theory, J. Masreliez shows that a "cosmic drag" force results from space and time expansion, which also predicts spiral star motions. In particular (Masreliez, 2012), he gets an additional radial acceleration term $-H\dot{r}$, with constant tangential velocity $v_0$ and exponential inward trajectories:

$$r \approx e^{-Ht} \text{ or } \frac{\dot{r}}{r} = -H \qquad (18)$$

In these conditions, our expansion acceleration is equivalent to his cosmic drag:

$$\frac{\dot{r}}{r}\dot{r} = -H\dot{r} \qquad (19)$$

The Masreliez theory also predicts the same radial acceleration:

$$2\dot{r}\dot{\theta} + r\ddot{\theta} = \frac{1}{r}\frac{d}{dt}(r^2\dot{\theta}) = \frac{1}{r}\frac{d}{dt}(v_0 r) = \frac{\dot{r}}{r}v_0 \qquad (20)$$

In this sense, acceleration (4) can be envisioned as a consequence of SEC theory. Nevertheless, our proposal postulates a non homogeneous expansion, depending on the scale level of the considered space, which leads to more precise descriptions, such as eq. (9) and other consequences developed in the next paragraphs.

Furthermore, contracted spiral trajectories have been assumed by J. Masreliez, but we have shown that an expanded solution can also be envisioned, depending on the sign of $\dot{r}$.

3.6 *Tully Fisher law*

Let us call:

$$\gamma_0 = \frac{v_0^2}{r_M} \qquad (21)$$

In the upper range of the galaxy ($r \cong r_M$), the expansion term can be neglected (eq. 14). Classical Newtonian gravitation can be applied here to the quasi circular trajectories:

$$\frac{v_0^2}{r_M} \cong \Gamma(r_M) \cong \frac{GM_t}{r_M^2} \qquad (22)$$

Eliminating $r_M$ from (21) leads to:

$$v_0^4 \cong \gamma_0 GM_t \qquad (23)$$

This expression relates the <u>constant</u> transverse velocity $v_0$ to the total mass, via $\gamma_0$ which is nothing else than the minimum central acceleration, obtained on the outskirts of the galaxy[1]. Since this minimum acceleration does not depend on the considered galaxy (McGaugh, 2011), we are then driven to postulate that there could be a <u>minimum (non zero) possible acceleration in the universe</u>. Galaxies could spread up to the radius corresponding to that minimum acceleration.

In this way and under our assumptions, we obtain a formal derivation of the Tully Fisher law. To be clear with our assumptions, it must be emphasized that the flat curves are explained by the additional cosmic acceleration. The TF law then results from the hypothesis of a minimum possible acceleration in the universe at a given time.

The experimental verifications of the Tully Fisher relationship constitute an experimental proof of the fact that $\gamma_0$ is a universal constant, and that the ratio

$$\frac{v_0^2}{r_M} \cong \frac{GM_t}{r_M^2}$$ does not depend on the considered galaxy.

3.7 *Towards a universally scaled space expansion*

The expansion coefficients obtained above are similar to the Hubble Constant, but with other values (eq. 12, 13, 15).

Concerning our galaxy, for a measured velocity $v_0$ of about 220 km/s and $\gamma_0 = 10^{-10} m/s^2$, eq (23) implies a mass $M_t = 3,5 \ 10^{41}$ kg. and $r_M$ can be obtained from (11) to be 16 kpc.

We obtain for $\omega_0$ (eq.13) the value 14 km/s/kpc or $4,6 \ 10^{-16} s^{-1}$.

Furthermore, the following expressions can be deduced from (13), and (21):

---

[1] Simple gravitation would predicts : $\gamma = \Gamma(r) = \frac{v_\theta^2}{r}$ and $v_\theta^2 = r\Gamma(r)$ should be radius dependant. The case $\Gamma(r) = \frac{GM(r)}{r^2}$ should predict $v_\theta^4 = \gamma GM$ where $v_\theta$, $M$ and $\gamma$ are radius dependant.

$$\gamma_0 = \omega_0 v_0 = \omega_0^2 r_M \qquad (24)$$

And from (22):

$$\omega_0^4 = \frac{\gamma_0^3}{GM_t} \qquad (25)$$

Eq (24) has the form of the famous Hubble expansion law, but with two differences:

$v_0$ is a transverse velocity and the value of $\omega_0$ depends on $\gamma_0$ and $GM_t$.

Space expansion coefficient depends on the « local » massive content.

From (11) and introducing the volumic mass density $\mu$, we obtain:

$$\omega_0 = \sqrt{\frac{4}{3}\pi G \mu} \qquad (26)$$

Expansion is proportional to the square root of the volumic mass density.

Inversely, expansion will decrease whenever the massive content vanishes.

3.8 *The universal constant $\gamma_0$ and the scaling of the expansion law*

The angular momentum can also be considered. Its maximum is (from (21) and (23):

$$C = v_0 r_M = \gamma_0^{-\frac{1}{4}} (GM_t)^{\frac{3}{4}} \qquad (27)$$

Let us try to transpose this result to the whole universe. We would obtain the following cosmological result, for a (2D) universe of radius R:

$$c^4 R^4 = \frac{(GM)^3}{\gamma_0} \qquad (28)$$

where M is the total mass of the universe and $c$ the velocity of light.

Allowing it to be 3 D, this can give an estimation of the universal constant $\gamma_0$, as a function of the mean massic (volumic) density $\mu_0$ of the universe. Using $R = \frac{c}{H}$, we obtain:

$$\gamma_0 = \left(\frac{4}{3}\pi G \mu_0\right)^3 \frac{c}{H^5} \qquad (29)$$

For the critical mass density $\mu_0 = \dfrac{3H^2}{8\pi G}$ we finally obtain:

$$\gamma_0 = \dfrac{Hc}{8} \tag{30}$$

This expression is intuitively significant to mean a small acceleration in the universe; and its numerical value is correct (0,86 $10^{-10}$ m/s² for H=2,310$^{-18}$ s$^{-1}$). It can also be considered as a derivation of the Hubble constant value:

$$H = 8\dfrac{\gamma_0}{c} \tag{30b}$$

which is numerically well verified for our galaxy.

These results validate our theory. Then Tully Fisher law (23) is better written as:

$$8v_0^4 \cong HcGM_t \tag{23b}$$

Furthermore, comparing (24) with (30) results in the scaling relationship:

$$\dfrac{H}{\omega_0} = 8\dfrac{v_0}{c} \tag{31}$$

Expansion seems to be <u>inversely proportional to velocity</u>.

3.9 *About Kepler laws for planet kinetics*

The well-verified Kepler laws do not obey eq (1b) since they require a constant angular momentum and not a constant transverse velocity.

However, let us compute the variations of the angular momentum C. We obtain:

$$\dfrac{dC}{dt} = v_0 \dot{r} \quad \text{and} \quad \dfrac{dC}{Cdt} = \dfrac{\dot{r}}{r} \tag{32}$$

We apply to the solar system the same model as the one developed for the galaxy, but consider that the planets are in free space, far away from the heart of the condensed matter (which is the sun). Then, in analogy with (9) and (15), we obtain:

$$\dfrac{\dot{r}}{r} \cong h \cong \dfrac{\omega_0}{\sqrt{3}} \tag{9b}$$

which represents the expansion coefficient for the solar system. It is reasonable to think that its value is similar to the one for the galaxy.[2] Then, from (32) and (9b), the order of magnitude of the angular momentum variation per year is:

$$\frac{\Delta C}{C} = h\frac{2\pi}{\dot{\theta}} = 2\pi \frac{4{,}6 \; 10^{-16}}{2 \; 10^{-7}} \cong 10^{-8} \qquad (33)$$

Where $\dot{\theta}$ represents here the earth angular velocity.

A waiting time of $10^7$ years is needed to obtain a significant change in angular momentum. For the planet dynamics in the solar system, the expansion acceleration is extremely weak when compared to the Newtonian one, since they are in the ratio (5):

$$\frac{\dot{r}}{v_\theta} = \frac{h}{\dot{\theta}}$$

This is why Kepler equations (1 and 2) remain valid for planetary motions, at least as an approximation for the orders of magnitude of our solar system.

[For the galaxy, the angular velocity is $10^9$ times much slower and the angular momentum variation (32) cannot be neglected any more].

As a summary, elliptic orbits occur in planetary systems where expansion is weak (eq 1 and 2). But at the galaxy level, eq. 1b and 2b allow spiral star orbits with a quasi constant transverse velocity.

Nevertheless, it must be pointed out in the context of our theory that the planetary orbits remain to be more precisely ruled by eq. (1b) and (2b), making them a good candidate to elucidate the problem of observed planet drifts.

**4 Conclusion and comments**

We have shown that the transverse velocity of planar mass distributions can be modelized by the use of purely Newtonian dynamics, under the new hypothesis of a universal space expansion acceleration proportional to velocity. Then, the Tully Fisher law has been derived, assuming a minimum universal acceleration in the universe. Tentative explanation of the fact that the velocity is not constant but is often slowly decreasing with $r$ could refer to other energy exchange processes, such as collisions which can be envisioned to take off energy whenever masses are not too close together, i.e. in the large radius range.

---

[2] Eq. (26) could be extrapolated, estimating the corresponding "local" volumic mass densities for the solar system ($M_S$ over a sphere of 4 light-years radius) and for the galaxy ($M_t$ over a sphere of 15 kpc radius) which gives the same order of magnitude.

Our study tends to argue that <u>extraneous hypotheses</u> (MOND theory or non baryonic dark matter) <u>are not necessary</u>, at least for this particular problem of galaxy velocities. Our conjecture of a <u>universally scaled expansion force</u> (eq. 4) has been introduced, resulting in a correct estimation of $\gamma_0$ (eq.30) and linking expansion to the local mass content (eq 26). This conjecture introduces a tentative hypothetical new fundamental dynamics principle, stating that the natural "free" movement would a priori be exponentially time-scheduled, thus asserting the Masreliez idea of time expansion. Applying it to the planet kinetics, we have shown that Kepler laws remain valid, at least as a convenient approximation for the orders of magnitude of the solar system.

It will be necessary to confront our conjecture with experiments and further theoretical developments.

Experimental observations of planetary drifts could be a first idea. More precise observations of the galactic variations of $v_\theta(r,t)$ and $\ddot{r}(r,t)$ would be also very interesting. Unfortunately, measurements are not easy, due in particular to the very low acceleration values. But indirect identifications could be fruitful: for instance, the expansion coefficient could be deduced from (26), the mass repartition parameters can be deduced from close observations of such curves as shown fig.1, etc. More simply, close observation of plain "linear" movement of "free" spacecrafts (Masreliez, 2005; Minguzzi, 2006; Nottale, 2003) could be of particular interest to verify such accelerations as stated by (4) or (30).

Further theoretical developments will also be useful to refine the mass density model and study the evolution with time, add a model of exchange processes between stars, generalize our formulation to 3D space, apply our proposal to other problems at various scales, extend the proposal to relativity and, in particular, link it with Nottale's ideas on scale relativity.

**Appendix: verification of the Euler equation**

For a velocity $v_\theta(r,t)$ (angular symmetry is assumed) we have:

$$\frac{dv_\theta}{dt} = \frac{\partial v_\theta}{\partial r}\dot{r} + \frac{\partial v_\theta}{\partial t} = 0$$

or (for $\dot{r} \neq 0$):

$$\frac{\partial v_\theta}{\partial r} = -\frac{\partial v_\theta / \partial t}{\dot{r}}$$

The radial dependance of the velocity is the ratio of its time evolution over the expansion rate. Postulating that the former is much lower that the latter leads to the result that the transverse velocity remains quasi constant with $r$.

More generally, we can verify the Euler equation:

$$\frac{d\vec{v}}{dt} = \frac{\partial \vec{v}}{\partial t} + rot\vec{v} \wedge \vec{v} + \frac{1}{2}grad\vec{v}^2 = \left(-\Gamma_t(r) + \frac{\dot{r}^2}{r^2}\right)\vec{i} + \frac{\dot{r}}{r}v_0\vec{j}$$

where $\vec{i}$ and $\vec{j}$ are the radial and transverse unitary vectors.

We postulate a constant transverse velocity $r\dot{\theta} = v_0$:

$$\vec{v} = \dot{r}(r,t)\vec{i} + v_0\vec{j}$$

We have:

$$\frac{\partial \vec{v}}{\partial t} = \frac{\partial \dot{r}}{\partial t}\vec{i}$$

and:

$$\frac{d\vec{v}}{dt} = (\ddot{r} - v_0\dot{\theta})\vec{i} + \dot{r}\dot{\theta}\vec{j}$$

$$rot\vec{v} = \frac{1}{r}\left(\frac{\partial}{\partial r}rv_0 - \frac{\partial \dot{r}}{\partial \theta}\right)\vec{k} = \frac{v_0}{r}\vec{k}$$

since $\dfrac{\partial \dot{r}}{\partial \theta} = 0$.

$[\vec{k} = \vec{i} \wedge \vec{j}\,]$

and

$$rot\vec{v} \wedge \vec{v} = -\frac{v_0^2}{r}\vec{i} + \frac{v_0\dot{r}}{r}\vec{j}$$

$$\frac{1}{2}grad\vec{v}^2 = \dot{r}grad(\dot{r}) = \dot{r}\frac{\partial \dot{r}}{\partial r}\vec{i}$$

Then the Euler equation is (for the only two gravitation and expansion forces):

$$\frac{d\vec{v}}{dt} = \frac{\partial \vec{v}}{\partial t} + rot\vec{v} \wedge \vec{v} + \frac{1}{2}grad\vec{v}^2 = \left(-\Gamma_t(r) + \frac{\dot{r}^2}{r^2}\right)\vec{i} + \frac{\dot{r}}{r}v_0\vec{j}$$

Assuming $\dfrac{\partial \dot{r}}{\partial \theta} = 0$ (axisymmetry), we obtain, along vectors $\vec{i}$ and $\vec{j}$ respectively, the following equalities:

$$\ddot{r} - v_0\dot{\theta} = \frac{\partial \dot{r}}{\partial t} - \frac{v_0^2}{r} + \dot{r}\frac{\partial \dot{r}}{\partial r} = -\Gamma_t(r) + \frac{\dot{r}^2}{r} \qquad (A1)$$

$$\dot{r}\dot{\theta} = \frac{v_0\dot{r}}{r} = \frac{\dot{r}}{r}v_0 \qquad (A2)$$

Knowing that $\ddot{r} = \dfrac{\partial \dot{r}}{\partial t} + \dot{r}\dfrac{\partial \dot{r}}{\partial r}$ and $r\dot{\theta} = v_0$, these equations are verified and eq.(A1) is equivalent to (2d). We have thus proved that a constant transverse velocity is possible, under the condition that acceleration (4) is added.

**Acknowledgments**

I am grateful to Dr. J. Masreliez who helped me to identify the link to his work and I also thank Dr. Michel Mizony and Dr. Robert Schmidt and for their help in the choice of a density repartition model.

**References**

Courteau, S., Andersen, D.R., Bershady, M.A., MacArthur L. A., Rix, H.W., C.R. 2003, Astrophys. J. 594, 208

Palunas, P., Williams, T.B., C.R. 2000, Astron. J. 120, 2884

McGaugh, S.S., C.R. 2011, Phys. Rev. Lett. 106, 303

Tonini, C., Maraston, C., Ziegler, B., Böhm, A., Thomas, D., Devriendt, J., Silk, J., C.R. 2011, MNRAS, 415, 811

Mo,H.J., Mao, S., C.R. 2000, MNRAS, 318, 163

Verheijen, M. A. W., C.R. 2001, Astrophys. J., 563, 694

McGaugh, S.S., C.R. 2002, Annu. Rev. Astron. Astrophys., 40, 263

Cardone, V.F., Angus, G., Diaferio, A., Tortora, C., Molinaro, R., C.R. 2011, MNRAS, 412, 2617

Bottema, R. Verheijen, M.A.W., 2001, Astron. and Astrophys., 388, 793

Bottema, R., C.R. 2002, Astronomy and Astrophys. 388, 809

Feng; J.Q., Gallo, C.F., C.R. 2010, J. Cosmology 6, 1373

Bienaymé, O., C.R. 1999, Astron. and Astrophys., 341, 86

Taylor, J.A., C.R. 1998, Astrophys. J. Lett. 497, 81

Mizony, M., 2003, « *La relativité générale aujourd'hui ou l'observateur oublié* », Aléas, Paris

Fuchs, B., Böhl, A., Möllenhoff, C., Ziegler; B.L., C.R. 2004, Astron. and Astrophys., 427, 95

Cooperstock, F.I., Tieu, S., C.R. 2007, Int. J. Mod. Phys., A22, 2293

Riess, A.G., C.R. 1998, Astron. J., 116, 1009

Nandra, R., Lasenby, A.N., Hobson, M.P., C.R. 2012, MNRAS, 422, 2945

Masreliez, J. C., 2012, "*The Progression of Time*", C. Johan Masreliez, Appendix III

Masreliez, J.C., C.R. 2004a, Apeiron 11, 99

Masreliez, J.C., C.R. 2004b, Apeiron 11, 1

J. Binney, J., Tremaine, S., 1994, "*Galactic Dynamics*", Princeton University Press, p. 43

Masreliez, J.C., C.R. 2005, Astrophys. and Space Science, 299, 1

Minguzzi, E., C.R. 2006, New Astron., 12, 142

Nottale, L., U. 2003, preprint, astro-ph.gr-qc/0307042v2